\documentclass[aps,prl,showpacs,twocolumn]{revtex4-1}        
\usepackage{epsfig}
\usepackage{dcolumn}
\usepackage{epstopdf}
\usepackage{graphicx}
\usepackage{color}

\begin{document}
\title{Low In solubility and band offsets in the small-$x$ $\beta$-Ga$_2$O$_3$/(Ga$_{1-x}$In$_x$)$_2$O$_3$ system}

\author{Maria Barbara Maccioni, Francesco Ricci, and Vincenzo Fiorentini}

\address{Department of Physics, University of Cagliari, and CNR-IOM, UOS Cagliari, Cittadella Universitaria, 09042 Monserrato (CA), Italy}
\begin{abstract}
Based on first-principles calculations, we show that the  maximum reachable concentration $x$  in the  (Ga$_{1-x}$In$_x$)$_2$O$_3$ alloy in the low-$x$ regime (i.e. In solubility in $\beta$-Ga$_2$O$_3$)  is  around  10\%.  We then calculate  the band alignment at the (100) interface between $\beta$-Ga$_2$O$_3$ and  (Ga$_{1-x}$In$_x$)$_2$O$_3$ at 12\%, the nearest computationally treatable concentration.
The alignment is strongly strain-dependent: it is of type-B staggered when the alloy is epitaxial on Ga$_2$O$_3$, and type-A straddling  in a free-standing  superlattice. 
Our results suggest a limited range of applicability of low-In-content GaInO alloys.
\end{abstract}

\maketitle
\noindent
The wide-band gap and large-breakdown-voltage insulator Ga$_2$O$_3$ is  attracting interest   for high-power transport, transparent electronics, and ultraviolet sensing applications. Combined with In$_2$O$_3$ (already widely used as transparent conducting oxide), Ga$_2$O$_3$   may originate a new (Ga$_{1-x}$In$_x$)$_2$O$_3$ materials system   enabling the band-engineering and nanostructuration concepts from popular semiconductor systems (such as, e.g., arsenides and nitrides) in a previously impervious region of high absorption energies and breakdown voltages. 
In this Letter we provide two key pieces of information for this endeavor, namely the maximum concentration of indium in the alloy and the interface band offset, which are hitherto unknown to our knowledge. 

We first address the degree of miscibility of Ga$_2$O$_3$ and In$_2$O$_3$. The parent materials
have different structures (monoclinic $\beta$ and cubic bixbyite, respectively), so the low-In and high-In-content alloying limits will be different, with likely complicated phase mixing at intermediate concentrations \cite{zang,ikz}.  Here we consider the alloying of $\beta$-Ga$_2$O$_3$ with In, and  show,  based on ab initio calculations, that In can be incorporated  into $\beta$-Ga$_2$O$_3$ at most at the 10\% level at typical growth temperatures. This agrees with the most recent estimate \cite{ikz} of around 10\%. We then address the  band offsets at the (100)  interface of $\beta$-Ga$_2$O$_3$ to the (Ga$_{1-x}$In$_x$)$_2$O$_3$ alloy, both epitaxial on Ga$_2$O$_3$ and free-standing. Given that $x$ is at most around 10\%, we study the offset in the computationally-affordable case of 12\% In. We find that the alignment is of type-B staggered when the  alloy is epitaxial on Ga$_2$O$_3$, and type-A straddling  in a free-standing  superlattice.


Alloying of monoclinic $\beta$-Ga$_2$O$_3$, the stable phase at ambient condition \cite{phases}, is simulated by substituting Ga with
  In  at various nominal concentrations and configurations. The interface is then simulated by a superlattice supercell. All optimizations (internal geometry, volume, etc.) and electronic structure calculations are done within  density functional theory (DFT) in the generalized gradient approximation (GGA), and the projector-augmented wave (PAW) method  as implemented in the VASP code  \cite{vasp}. The PAWs include occupied $d$ states in the valence for both cations. For the alloy calculation we use an 80-atom (32-cation)  supercell  containing 1$\times$4$\times$1 20-atom conventional cells, and for the interface calculation a 160-atom (32-cation) 2$\times$2$\times$2 supercell. The k-point sampling is on a  2$\times$4$\times$2  grid. We work at the calculated lattice parameters $a$=12.46  \AA, $b$=3.08 \AA, $c$=5.88 \AA, $\theta$=103.65$^\circ $, which compare well with experiment  \cite{geller,mbm}.

We choose as dilute limit the concentration  of   3\% In, i.e. one ``isolated" In atom per 80-atom cell.  Besides being  computational feasible,  3\%   is actually a quantitatively accurate dilute limit:  the formation energy calculated in the standard way \cite{vdw} is $E_f(1)$=0.24 eV/In, which yields a concentration of 2.7\%  at the typical growth temperature $T_g$=775$\div$800 K \cite{zang,ikz}. The chemical-potential reservoir for In is the bixbyite phase of In$_2$O$_3$, which  might occur in  nanograins  embedded in Ga$_2$O$_3$.

\begin{figure}[h]
\begin{center}
\includegraphics[width=8cm]{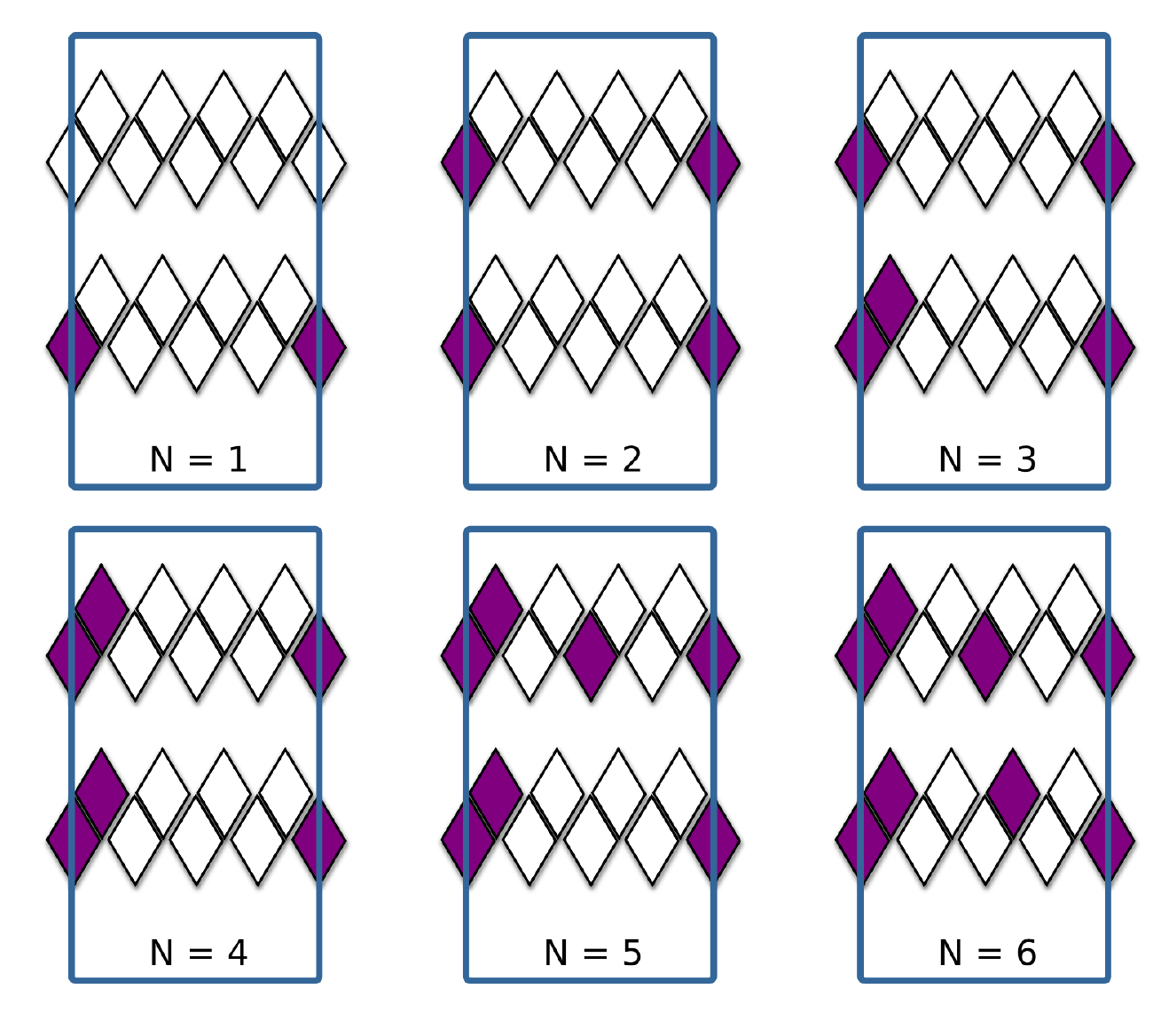}
\end{center}
\caption{\label{str} (Color online) Sketch of different configurations of In on the Ga$_2$O$_3$  simulation supercell. Occupied octahedra in the $\beta$ structure double-rows are darkened.}
\end{figure}

Indium substitution at tetrahedral sites costs $\delta E_t$=1.1 eV more than at octahedral sites; thus the tetrahedral-site  occupation probability is lower than that of octahedral sites by a factor $\exp{(-\delta E_t/k_BT_g)}$$\sim$0.5-1$\times$10$^{-7}$. Therefore, the  In concentration in Ga$_2$O$_3$  cannot exceed the value whereby the octahedral  sites are all occupied, i.e. 50\%. 
Because the Ga$_2$O$_3$ structure is made up of double rows of octahedra sharing sides and  connected by tetrahedra,  there is limited configurational lee-way for  In placement in the system (see Fig.\ref{str}; for a more realistic depiction see e.g. Ref.\cite{mbm}). We evaluate the energetics of In substitution in various  configurations (a sample is depicted in  Fig.\ref{str}) at concentrations between 6\% and 25\%, i.e. for 2 to 8 In atoms in the 80-atom, 32-cation  1$\times$4$\times$1 cell, and extrapolate numerically to 16 atoms per cell (tetrahedral sites are neglected).
 We find that two In's prefer to sit  on different double-rows or, failing that (as inevitably is the case for growing $x$), on   first-neighbor octahedra in adjacent subrows, which locally resemble the native In$_2$O$_3$ bixbyite structure. The formation energy per In decreases slightly for two and three In per cell, then increases steadily.  
 For  the configurations in Fig.\ref{str} we find that the excess formation energies over that of a single In are $\delta E_f(2)$=--0.044, $\delta E_f(3)$=--0.019, $\delta E_f(4)$=+0.021, $\delta E_f(5)$=+0.074, $\delta E_f(6)$=+0.144, $\delta E_f(7)$=+0.171, $\delta E_f(8)$=+0.180, in eV/In (the last two are not shown in the Figure).
The cell is kept at the volume of the undoped material, which is strictly correct in the dilute limit \cite{finnis}; at higher concentration we account for an enthalpic energy cost (see below).  The concentration is evaluated as the thermal average of the In population in the supercell ($M$=32 cation sites)
\begin{equation}
x=\frac{\langle N\rangle}{M}=\frac{1}{M}
\frac{\sum_{N=1}^{M} N \exp{[-\beta_gF(N)]}}{\sum_{N=1}^{M}\exp{[-\beta_gF(N)]}}, 
\label{aver}
\end{equation}
where $\beta_g$=1/$k_B$$T$$_g$ and $F(N)$=$E_f(1)$+$\delta E_f(N)$--$T_gS$+$\delta H$ is the free energy per In in the $N$-In substituted cell.  $E$ is the formation energy, $S$ the formation vibrational  entropy (we estimate it from the Debye temperature of the two bulk oxides, and find $T_gS$$\simeq$0.015 eV), and $\delta H$$\simeq$0.09 eV is the energy cost related to the  internal pressure building up in the constrained cell. $\delta H$ is estimated as the energy difference (per In) between the constrained and volume-relaxed  cell; if cell-length changes are allowed along a given direction, as would occur in epitaxy,  $\delta H$ decreases by about one third. In any event, as we have seen, entropy and enthalpy provide only small corrections over the structural energy  $E_f$ discussed above.
 The  thermal population average, Eq.\ref{aver},  gives a concentration of 
 9\%, with an error bar of +2\%  and --1\%  estimated  varying the $\delta E$'s between 0.5 and 1.5 times those calculated. Again, this low solubility follows from tetrahedral sites being ruled out and from In occupying only about 3 out of 16 octahedral sites in the cell on (thermal) average.

\begin{figure}[h]
\begin{center}
\includegraphics[width=7.5cm]{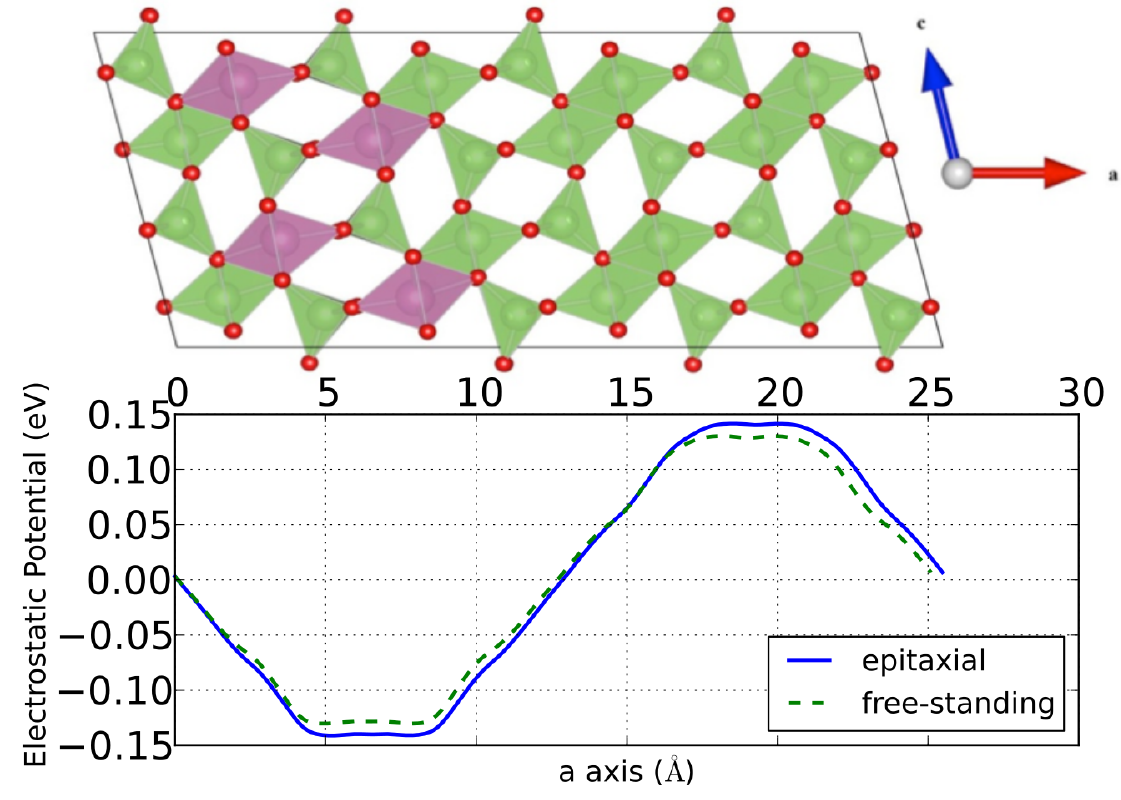}
\end{center}
\caption{\label{cell} (Color online) Upper panel: simulation cell for the (100) superlattice (for definiteness we display the epitaxial geometry). Lower panel: the electrostatic potential of the superlattice, showing small but definite bulk regions on either side of the interface. The potential is aligned with the lower side of the cell.}
\end{figure}

Having established the small solubility of In in Ga$_2$O$_3$, we come to the band offsets. The correct way of calculating band offsets \cite{peressi} is as the sum $\Delta$E$_b$+$\Delta$$V$ of the interface jump $\Delta$$V$ in electrostatic potential between the two regions being interfaced, and the difference $\Delta$$E_b$ of the band edge of interest in each of the two materials, taken separately each in their own internal potential. As mentioned, we use a 2$\times$2$\times$2 160-atom cell, depicted in Fig.\ref{cell}, upper panel, to describe  the  (100) interface: half of the supercell along the (100) axis is pure Ga$_2$O$_3$, and the other half is a Ga-In alloy. We pick the concentration of 12\% as it is near the maximum  achievable (as discussed previously), and because, given the energetics constraints, the 
 configurational freedom of In is very limited, and there is no serious need for a  detailed In  configurations sampling, which would be computationally unfeasible. We choose the (100) interface for computational convenience; it remains to be assessed how much the offsets change with orientation.  

This super-unit cell  repeats periodically the two layers,  effectively  producing a superlattice;  we find that the thickness of the layers is sufficient to reproduce identifiable bulk regions on either side of interface, with flat, bulk-like  average potential, as shown in Fig.\ref{cell}, lower panel.
We study this  superlattice in two strain states, epitaxial and free-standing; in the former case we fix the lattice constants in the $b$-$c$ crystal plane and the monoclinic angle to those of Ga$_2$O$_3$, and relax the $a$  lattice parameter; in the second case, we optimize all lattice parameters. The internal coordinates are optimized in all cases. 

\begin{figure}[h]
\begin{center}
\includegraphics[width=8cm]{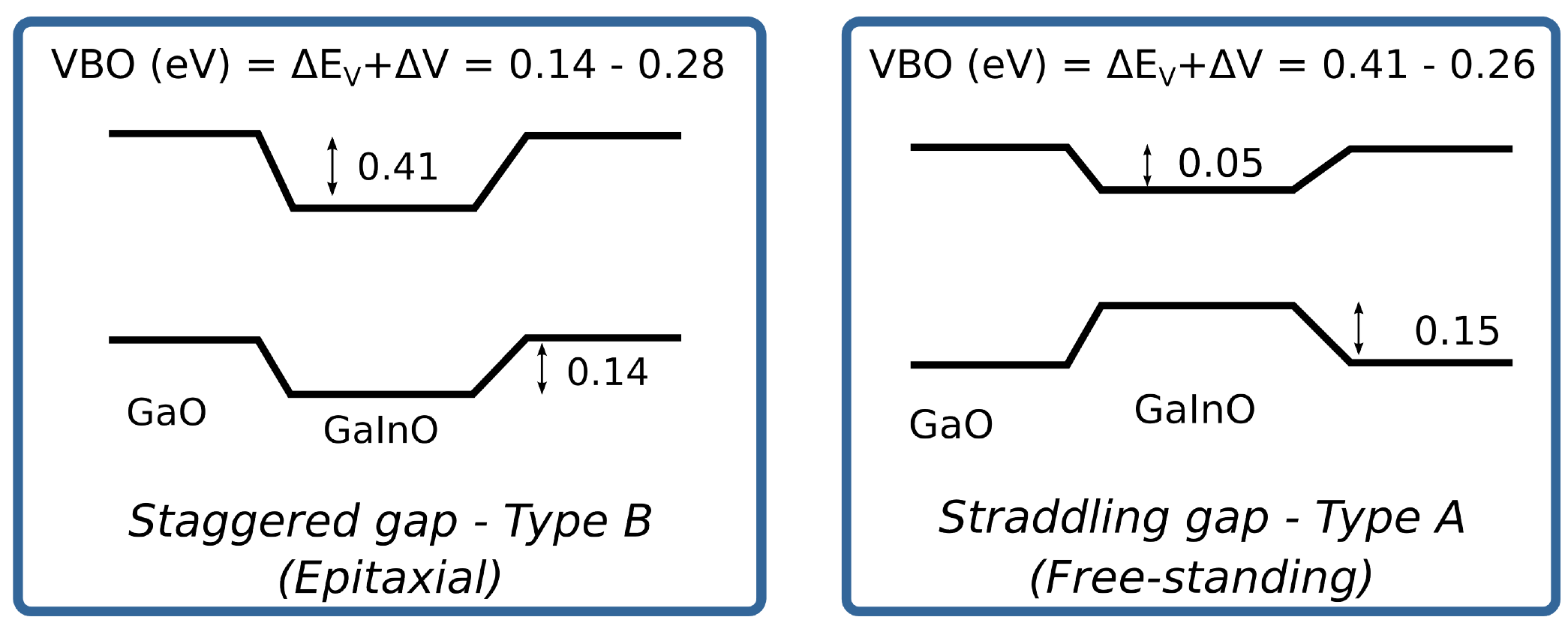}
\end{center}
\caption{\label{offs} (Color online) Schematic of the staggered and straddling offset for, respectively, the  epitaxial and free-standing  superlattice configurations.}
\end{figure}

As schematized in Fig.\ref{offs},   at the (100) interface between Ga oxide and the alloy at  12\% In, we find an alignment of type-B staggered when the alloy is epitaxial on Ga$_2$O$_3$, and type-A straddling  in a free-standing  superlattice; the valence offsets  from Ga$_2$O$_3$ to (Ga$_{1-x}$In$_x$)$_2$O$_3$ are --0.14 eV  (Ga$_2$O$_3$-epitaxial) and 0.15 eV (free-standing), and the conduction offsets are  \mbox{--0.41 eV} (epitaxial) and \mbox{--0.05 eV} (free-standing). This considerable difference is due almost entirely to strain-induced shifts of the  valence band maximum (VBM) and conduction band minimum (CBM), whereas the electrostatic interface alignment is hardly insensitive to strain. This indicates that a marked dependence on the strain state, and hence on the growth quality, is to be expected. Importantly, given the limited In solubility, this is about as much of an offset as can be expected between Ga$_2$O$_3$ and  (Ga$_{1-x}$In$_x$)$_2$O$_3$. There seems to be no measurement of the quantities just discussed, and we hope our prediction will stimulate work in this direction.

We expect the above estimate to be rather accurate. Our interface is between materials differing only very slightly due to compositional changes, so that beyond-DFT  corrections to the band edges will essentially cancel out; on the other hand, strain-induced  band-edge shifts are known to be well described by standard functionals \cite{vf92}. By the same token, in this case, the gap error also essentially cancels out, so the absolute value of the gap is immaterial to the offsets.
For completeness, we mention that the GGA gap is about 2 eV, i.e., as expected, a 60\% underestimate compared to experiment \cite{zang,ricci}. Adding an empirical self-energy correction \cite{fiore} involving the calculated high-frequency dielectric constant, we obtain a gap of 4.2 eV, not far from the most recent experimental and  theoretical beyond-DFT estimates  of 4.6 and 4.7 eV,  respectively, to be discussed elsewhere \cite{ricci}.  As  reported previously \cite{mbm}, the gap rates of change with composition and volume are also close to experiment \cite{zang}. 

In summary, we have performed first-principles calculations on  the bulk and interface properties of the Ga$_2$O$_3$/ (Ga$_{1-x}$In$_x$) system. Importantly, we find that  In is soluble in  Ga$_2$O$_3$ only up to a maximum of about 10\%. The band offset between Ga oxide and the alloy at  12\% In is of type-B staggered when the alloy is epitaxial on Ga$_2$O$_3$, and type-A straddling  in a free-standing  superlattice. The valence offsets  from Ga$_2$O$_3$ to (Ga$_{1-x}$In$_x$)$_2$O$_3$ are --0.14 eV  (Ga$_2$O$_3$-epitaxial) and 0.15 eV (free-standing), and the conduction offsets are  --0.41 eV (epitaxial) and --0.05 eV (free-standing).

Work supported in part by MIUR-PRIN 2010 project {\it Oxide}, Fondazione Banco di Sardegna and CINECA grants. MBM acknowledges  financial support of her PhD scholarship by the Sardinian Regional Government (P.O.R. Sardegna F.S.E. Operational Programme of the Autonomous Region of Sardinia, European Social Fund 2007-2013, Axis IV Human Resources, Objective l.3, Line of Activity l.3.1).

\end{document}